\documentclass[final]{aipproc}

\layoutstyle{6x9}

\begin{document}

\title{Modeling X-ray emission from stellar coronae}


\author{S. G. Gregory}{
  address={SUPA, School of Physics and Astronomy, Univ. of St Andrews, St Andrews, KY16 9SS, UK},
  ,email={sg64@st-andrews.ac.uk},
}
\author{M. Jardine}{
  address={SUPA, School of Physics and Astronomy, Univ. of St Andrews, St Andrews, KY16 9SS, UK}
}
\author{C.  Argiroffi}{
  address={Dipt. di Scienze Fisiche ed Astronomiche, Sezione di Astronomia, Univ. di Palermo, Piazza del Parlamento 1, 90134 Palermo, Italy},
  ,altaddress={INAF - Osservatorio Astronomico di Palermo, Piazza del Parlamento 1, 90134 Palermo, Italy}
}
\author{J.-F.  Donati}{
  address={LATT - CNRS/Universit\'{e} de Toulouse, 14 Av. E. Belin, F-31400 Toulouse, France}
}


\begin{abstract}
By extrapolating from observationally derived surface magnetograms of low-mass stars we construct models of their
coronal magnetic fields and compare the 3D field geometry with axial multipoles. AB~Dor, which has a radiative core, has a very
complex field, whereas V374~Peg, which is completely convective, has a simple dipolar field. We calculate global X-ray emission
measures assuming that the plasma trapped along the coronal loops is in hydrostatic equilibrium and compare the differences
between assuming isothermal coronae, or by considering a loop temperature profiles.  Our preliminary results suggest that the 
non-isothermal model works well for the complex field of AB~Dor, but not for the simple field of V374~Peg.
\end{abstract}

\maketitle


\section{Detecting and extrapolating stellar fields}
Using the spectropolarimetric technique known as Zeeman-Doppler imaging (ZDI) 
it is possible to map the medium and large scale structure of stellar magnetic topologies (see Donati et al, 
these proceedings).  From such observationally derived surface maps of the photospheric magnetic 
fields of the rapidly rotating K-dwarf AB~Dor \cite{don97} and the low mass M-dwarf
V374~Peg \cite{don06} we extrapolate the coronal fields assuming that they are potential, 
or current free, $\nabla \times \mathbf{B} = \mathbf{0}$ (Fig. \ref{extraps}).  This condition
is satisfied by writing the field in terms of a scalar flux function $\Psi$, with 
$\mathbf{B} =-\nabla \Psi$.  As the field must also be solenoidal, $\nabla \cdot \mathbf{B}=0$,
then $\Psi$ must satisfy Laplace's equation $\nabla^2 \Psi=0$, the solution of which is a linear
combination of spherical harmonics.  Thus the three components of the magnetic field vector
can be determined within a defined 3D volume, and a field line tracing algorithm employed 
to derive the coronal magnetic field topology.  The extrapolation technique is described 
in greater detail by \cite{jar02} and \cite{gre06a}.  The stellar parameters for AB~Dor 
and V374~Peg are listed in Table \ref{table_params}.


\subsection{Comparison with axial multipoles}
\begin{figure*}
        \centering
        \begin{tabular}{cc}
                 \label{extrap_abdor}      
                 \includegraphics[height=.30\textheight]{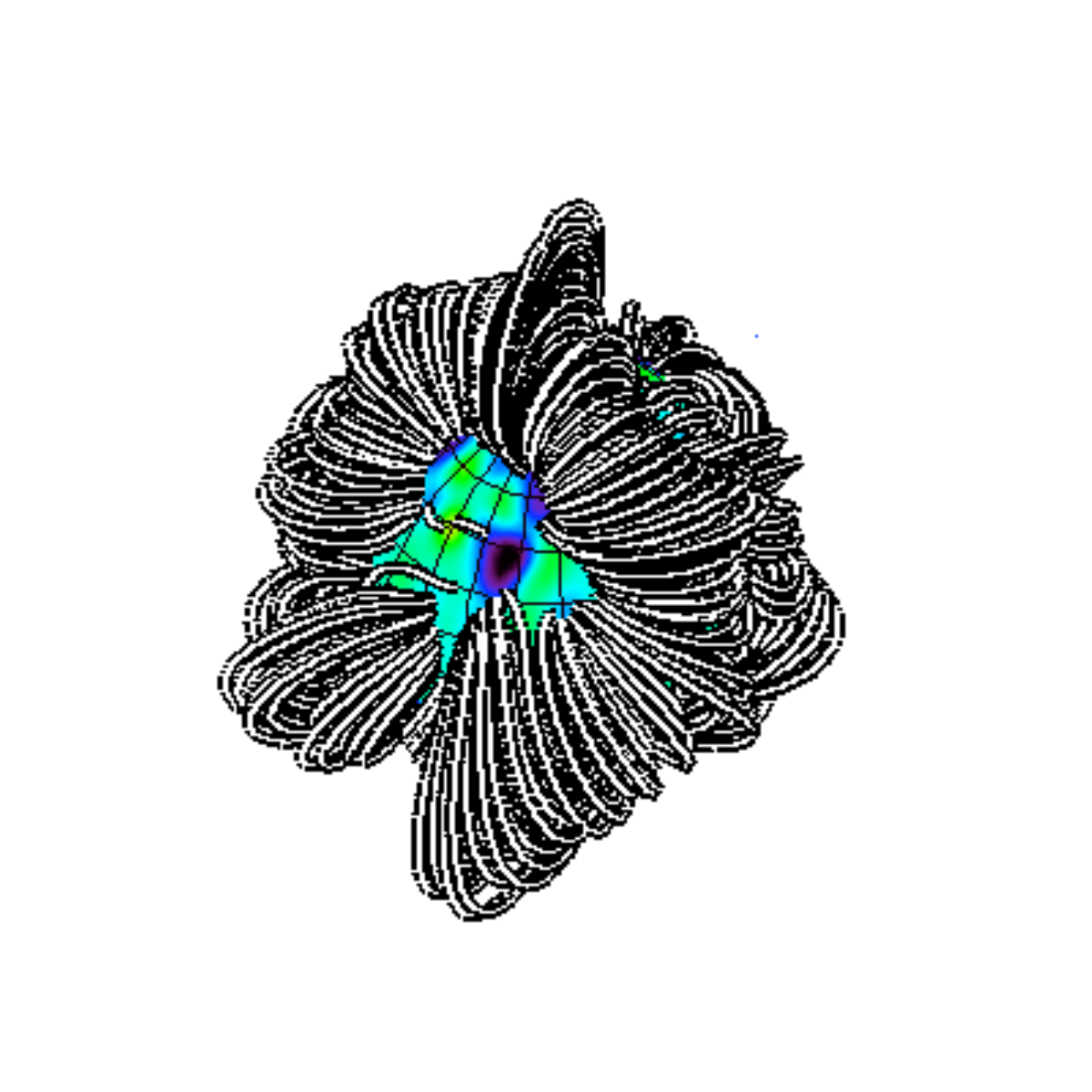}
                        &
                 \label{extrap_v374peg}
                 \includegraphics[height=.24\textheight]{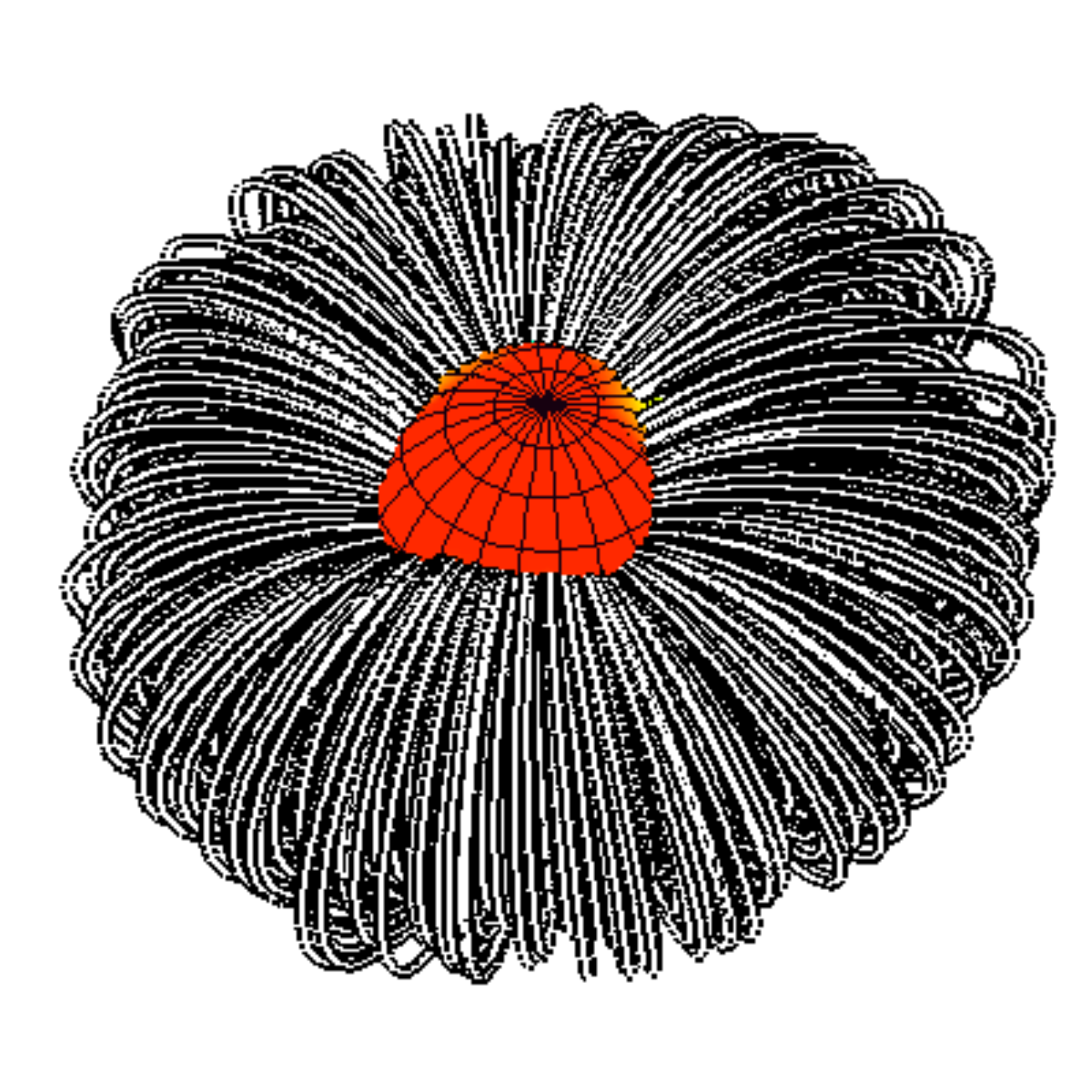}
                          \\                 
                 \label{hw_abdor}      
                 \includegraphics[height=.29\textheight]{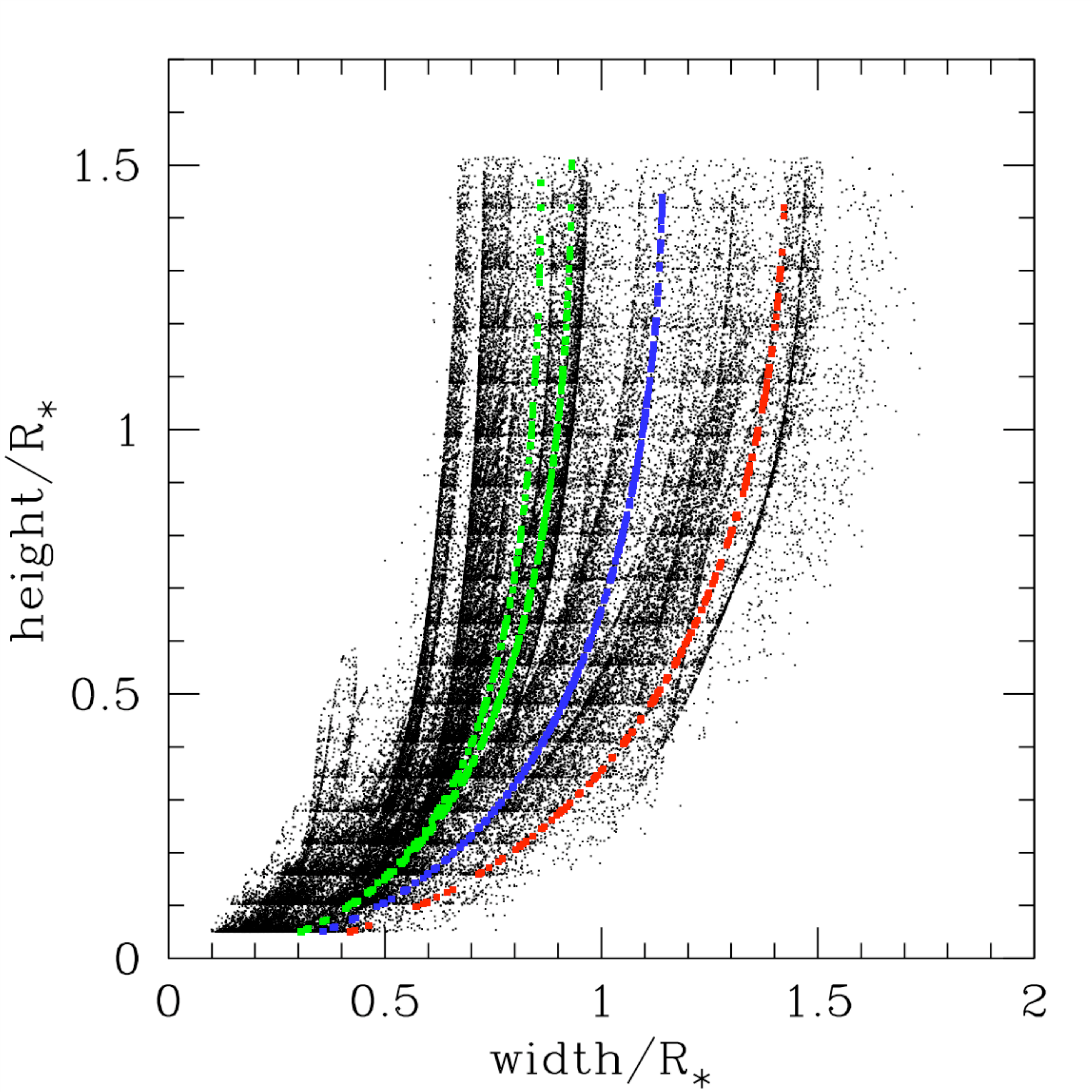}
                        &
                 \label{hw_v374peg}
                 \includegraphics[height=.29\textheight]{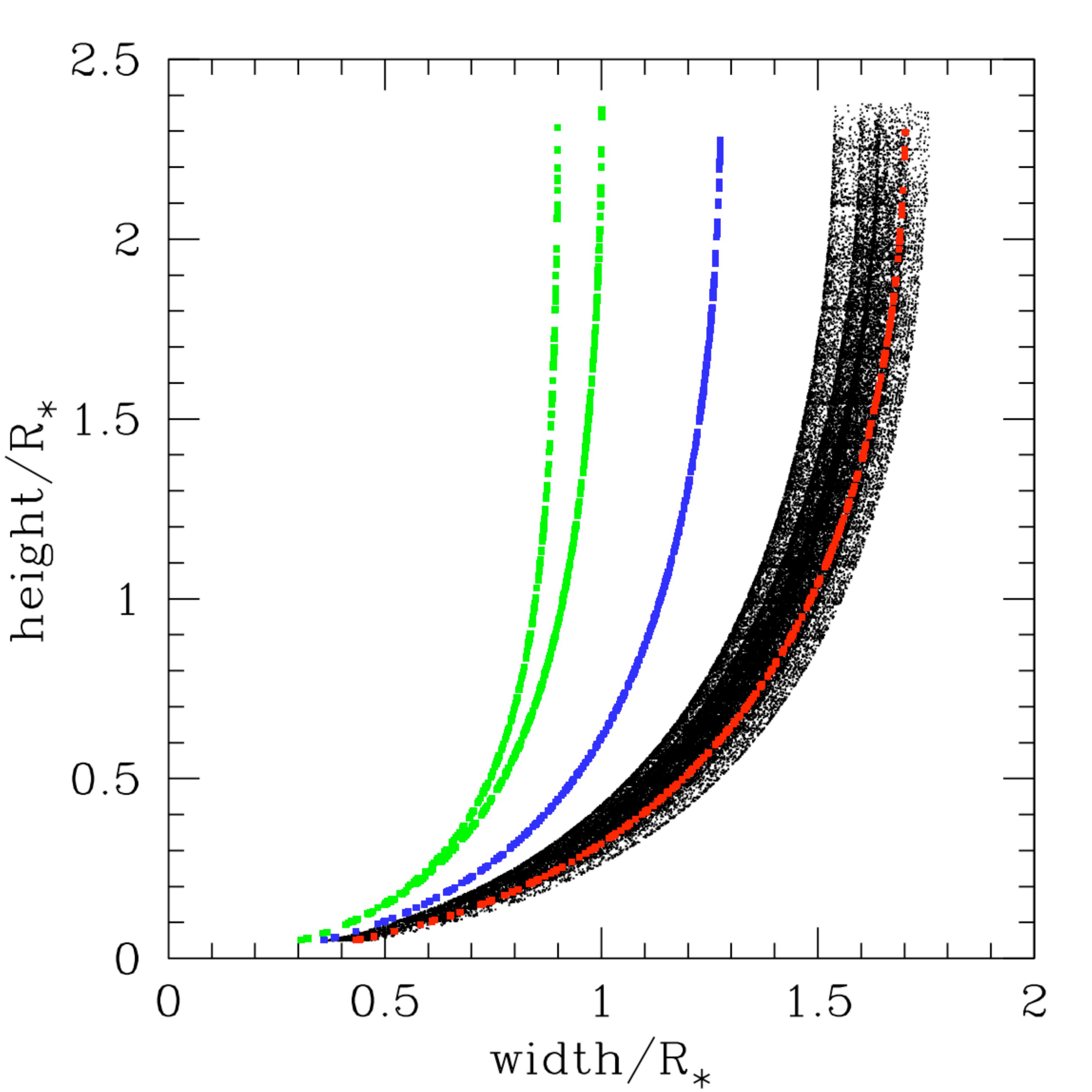}
                          \\
        \end{tabular}
        \caption{Field extrapolations from magnetic surface maps of AB~Dor (left) and V374~Peg (right)
                [upper panel, not to scale]. The plots [lower panel] show field line height vs width, as defined in Fig \ref{hw},
                compared to axial multipoles, a dipole (red), a quadrupole (blue) and an octupole (green).}
        \label{extraps}
\end{figure*}

\begin{figure}
  \includegraphics[height=.20\textheight]{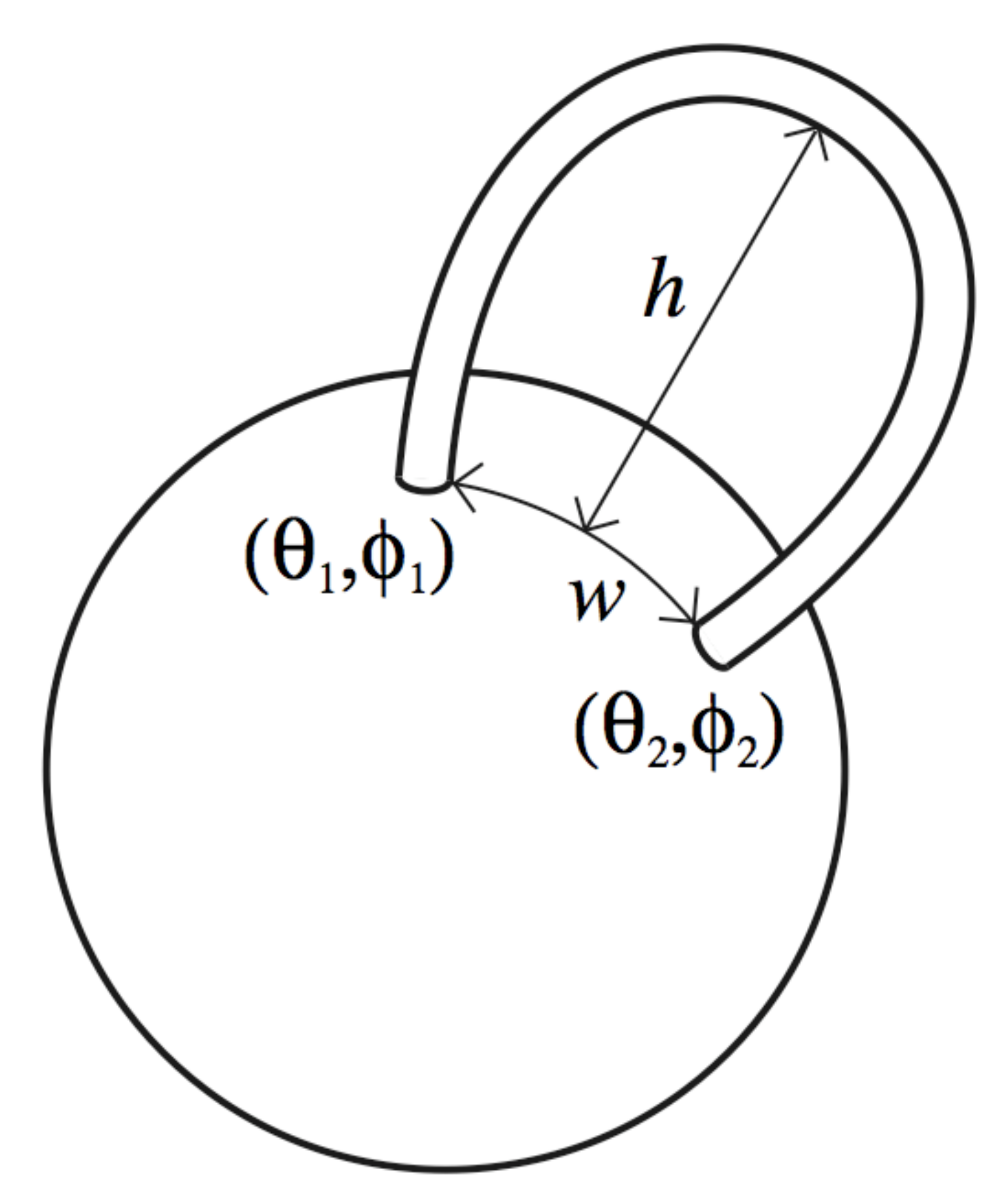}
  \caption{The width, $w$, of field lines is defined as the distance measured
           along the segment of the great circle connecting the field line foot points. It is
           calculated using the Haversine formula (equation \ref{haversine}).  The height, $h$, is the 
           maximum loop height above the stellar surface.}
  \label{hw}
\end{figure}

We compare the 3D coronal field structures obtained via field extrapolation 
with axial multipoles (Fig. \ref{extraps}). The height and width
of field line loops is defined in Fig. \ref{hw}.  The width is calculated using the Haversine 
formula,
\begin{equation}
w = 2\sin^{-1}{\left[\sqrt{\sin^2{\left(\frac{\Delta \theta}{2}\right)}+\sin{\theta_1}\sin{\theta_2}\sin^2{\left(\frac{\Delta \phi}{2}\right)}}\right]},
\label{haversine}
\end{equation}  
where $\Delta \theta =|\theta_2 - \theta_1|$ and $\Delta \phi = \phi_2 - \phi_1$ are the differences in co-latitudes
and longitudes of the field line foot points, and where all angles are measured in radians and all distances in units of 
stellar radii.  

V374~Peg, at only $0.3\,{\rm M}_{\odot}$, is completely convective \cite{don06} and has a large scale axisymmetric magnetic
field which has been found to be remarkably stable over a timescale of at least one year \cite{mor08}.  Its field is almost dipolar, as 
evident from both the field extrapolation and from the plot of field line height vs width (Fig. \ref{extraps}). 
In contrast, AB~Dor is a solar mass star with a radiative core, the star spot distribution (and therefore the magnetic field) on which is
known to vary on a timescale of at least one year \cite{jef07}.  The magnetic field of AB~Dor is particularly complex (Fig. \ref{extraps}) 
with contributions from many high order multipole components.  The difference in field complexity for both stars is likely to be a simple reflection of
their different internal structures, and therefore magnetic field generation mechanisms.  


\subsection{Coronal X-ray emission} 
We assume that the X-ray emitting plasma is trapped along the coronal loops and calculate the pressure
structure along loops by integrating the equation of hydrostatic equilibrium \cite{gre06b}.  The pressure at a point
$s$ along a loop is given by,
\begin{equation}
p(s) = p_0 \exp{\left[ \frac{\mu m_H}{k_B}\int_0^s \frac{g_s}{T} ds \right]},
\label{hydro}
\end{equation} 
where $g_s$ is the component of the effective gravity (i.e. the component of the combined centrifugal and gravitational
acceleration) along the direction of the field at $s$, and $g_s = \mathbf{g}\cdot\mathbf{B}/|\mathbf{B}|$.  Assuming that
the pressure and density are related via $p=2nk_BT$ then the global X-ray emission measure can be calculated
as described by \cite{gre06b}. We set the coronal pressure 
to zero for open field lines and for loops which would be unable to contain the coronal plasma due to the gas pressure
exceeding the magnetic pressure. Such regions therefore do not contribute to the X-ray emission.  X-ray emitting gas is 
typically contained within compact magnetic regions, however some regions of the stellar surface
remain dark in X-rays, giving rise to modulation of X-ray emission, especially for
AB~Dor. 

Table \ref{table} shows the simulated X-ray emission properties when assuming
an isothermal corona (at 10$\,{\rm MK}$ for AB~Dor and 20$\,{\rm MK}$ for V374~Peg).  Also shown for comparison are the 
same quantities derived by considering a Serio-type temperature scaling law \cite{ser81}, where the loop 
temperatures depend on the pressure at the loop foot points, the length of the loop, and a (somewhat arbitrary) heating
scale length.  By comparing our simulated X-ray luminosities with those derived from ROSAT data (Table \ref{table}), our preliminary results 
suggest that such a non-isothermal coronal model works reasonably well for AB~Dor with its complex magnetic structure, but fails 
for the simple field of V374~Peg.   

\begin{table}
\begin{tabular}{ccccc}
\hline
  \tablehead{1}{c}{b}{Star \\}
 & \tablehead{1}{c}{b}{Mass\\(M$_{\odot}$)}
  & \tablehead{1}{c}{b}{Radius\\(R$_{\odot}$)}
  & \tablehead{1}{c}{b}{P$_{rot}$\\(d)}
  & \tablehead{1}{c}{b}{d\\(pc)} \\
\hline
AB~Dor   & 1.0   & 1.0  & 0.51 & 14.94 \\
V374~Peg & 0.3   & 0.35 & 0.45 & 8.963 \\
\hline
\end{tabular}
\caption{Stellar parameters for V374~Peg and AB~Dor.}
\label{table_params}
\end{table}

\begin{table}
\begin{tabular}{ccccccc}
\hline
  \tablehead{1}{c}{b}{Star\\ \\}
  & \tablehead{1}{c}{b}{T$_{corona}$\\$[$iso$]$\\(MK)}
  & \tablehead{1}{c}{b}{log EM\\$[$iso$]$\\(cm$^{-3}$)}
  & \tablehead{1}{c}{b}{log EM\\$[$non-iso$]$\\(cm$^{-3}$)}
  & \tablehead{1}{c}{b}{log L$_X$\\$[$iso$]$\\(erg s$^{-1}$)} 
  & \tablehead{1}{c}{b}{log L$_X$\\$[$non-iso$]$\\(erg s$^{-1}$)}
  & \tablehead{1}{c}{b}{log L$_X$\\$[$obs, ROSAT$]$\\(erg s$^{-1}$)} \\
\hline
AB~Dor   & 10   & 52.48 & 52.75 & 29.69 & 29.94 & 30.18 \\
V374~Peg & 20   & 52.13 & 52.71 & 29.16 & 29.91 & 28.88 \\
\hline
\end{tabular}
\caption{Simulated/observed X-ray properties for V374~Peg and AB~Dor.}
\label{table}
\end{table}


\section{Future work}
We are currently adopting the models discussed by \cite{col88} and \cite{unr97}, in order to incorporate proper
temperature profiles along coronal loops (Gregory et al, in preparation).  Emission measures distributions 
(EMDs) will be derived and compared to those obtained from the observed X-ray spectra, allowing
the validity of temperature scaling laws to be examined when used in conjunction with extrapolated
magnetic fields.  For such field structures, the loop geometry often departs from the semi-circular
shape assumed by many previous studies (e.g. \cite{ser81}).  We will determine whether or not the same temperature
scaling laws can be applied to both stars with complex magnetic fields (like AB~Dor) and those with simple 
large-scale axisymmetric fields (like V374~Peg).     


\bibliographystyle{aipprocl}

\end{document}